\documentstyle[12pt]{ioplppt}

\newcommand{\r}{\rightarrow}
\newcommand{\BFACE}[1] {\mbox{\boldmath $#1$} }

\begin{document}
\jl{1}

\title{Non-integrability of the mixmaster universe}

\author{F Christiansen\dag, H H Rugh\ddag and S E Rugh\P}

\address{\dag\ Istituto Nazionale di Ottica,
Largo E. Fermi 6, Firenze, I-50125, Italy}

\address{\ddag\ Institut Hautes Etudes Scientifiques,
35 Route de Chartres, Bures-sur-Yvette, France}

\address{\P\ The Niels Bohr Institute,
University of Copenhagen,
Blegdamsvej 17, 2100 K\o benhavn \O, Denmark}

\begin{abstract}
\noindent
We comment on an analysis by Contopoulos et al. which demonstrates
that the governing six-dimensional
Einstein equations for the mixmaster space-time metric pass the ARS or
reduced Painlev\'{e} test. We note that this is the case irrespective
of the value, $I$, of the generating Hamiltonian which is a constant of
motion. For $I < 0$ we find numerous closed orbits with two unstable
eigenvalues strongly indicating that there cannot exist two
additional first integrals apart from the Hamiltonian and thus that the
system, at least for this case, is very likely not integrable. In
addition, we present numerical evidence that the average Lyapunov
exponent nevertheless vanishes.

The model is thus a very interesting example of a Hamiltonian dynamical
system, which is likely non-integrable yet passes the reduced
Painlev\'{e} test.

\end{abstract}

\pacs{0230Hq, 0420Dw, 0420Ha, 0425Dm, 0440Nr, 9880Hw}

\maketitle

\section{Introduction.}

In general, it is a difficult task to establish whether or not a
Hamiltonian system of $N$ degrees of freedom is integrable. We note
that there are many definitions of integrability \cite{Kruskal}. Here,
we take as definition~\cite{Arnold} the existence of $N$
{\em independent} first integrals $I_j = I_j (\BFACE{p}, \BFACE{q}) =
const $ which are {\em in involution}, i.e. their mutual Poisson
brackets vanish $\left\{ I_i, I_j \right\} = 0$, for all $i,j$. In case
that $N$ such integrals have not been found, one may try to establish
the integrability of the dynamical system via the singularity analysis
called the Painlev\'{e} test. According to the ``Painlev\'{e}
conjecture'' a Hamiltonian dynamical system is integrable if it has the
Painlev\'{e} property. However, since carrying out the Painlev\'{e}
test is often too difficult a task, in practice the test is performed
in a reduced form called the ARS-algorithm
\cite{Ablowitz77,Ablowitz80,Ramani89}. In this paper we shall comment
on an example where the (ARS) conjecture fails.

In Contopoulos et al. \cite{Contopoulos}, and in Cotsakis et al.
\cite{Cotsakis}, it is shown that the Einstein equations for the
mixmaster space-time metric pass the ARS-test (the reduced Painlev\'{e}
property) and, by the above proposition, that it is probably integrable.

Here we shall argue that this is very likely {\em not the case}. Our
argument pursues the following line: First, there is a generating
Hamiltonian, the value of which, $I$, characterizes different
qualitative behaviours of the model. The case $I=0$ corresponds to the
proper mixmaster universe. Although it contains the chaotic Gauss-map
in its time-evolution there is no direct contradiction in calling the
system integrable, e.g. in the sense of being solvable by an inverse
scattering transform (see e.g. \cite{Ablowitz77}). For $I>0$, the
evolution is in some sense `trivial', but for $I<0$\footnote{
  At first sight a universe model with $I<0$ might appear quite
  artificial. It has, however, a physical interpretation as a
  mixmaster universe in which matter with negative energy density
  is uniformly distributed. It gives the possibility of an oscillatory
  evolution of the three-volume of the universe \cite{RughM,RughJ} and
  thus the possibility of periodic (cyclic) mixmaster universes.
}
we shall show that the system displays chaotic behaviour in the sense
that it contains an abundance of unstable periodic orbits, even
apparently infinite families of these,
which in turn renders the existence of a full set of first integrals
very unlikely. In fact, the proliferation of unstable periodic orbits
indicates the existence of a `Smale horseshoe' prohibiting
integrability. A rigorous proof would require finding a transversal
homoclinic intersection \cite{Cushman,Niko84},
i.e. a transversal intersection of the stable and unstable
manifolds of a hyperbolic closed orbit of the system, a work
currently in progress.
Second, the Painlev\'{e} analysis for the mixmaster model as presented
in Contopoulos et al. \cite{Contopoulos} {\em does not} distinguish
the value of this first integral, i.e. by this analysis one arrives at
the contradictory statement that even in the chaotic case $I<0$ the
model is probably integrable\footnote{
   In a subsequent paper \cite{Contopoulos94ny} Contopoulos et al.
   cast doubt as to the conclusion of their original
   paper \cite{Contopoulos}. They introduce a {\em perturbative}
   Painlev\'{e} test which the mixmaster model fails to pass.
   Also, Latifi et al. \cite{Latifi94} note that the model does not
   pass the full Painlev\'{e} test.
}.

Finally, we present numerical evidence that, despite the existence of
unstable periodic orbits, the average Lyapunov exponent vanishes in the
case $I<0$, contradicting data based on single trajectories
presented in Rugh~\cite{RughM} and Hobill et al.~\cite{Hobill91}.

\section{The equations of motion.}

The mixmaster space-time metric is a famous cosmological toy-model
\cite{MTWLL} which has been studied extensively in the past two
decades in various contexts. It is an anisotropic generalization of the
standard cosmological model of our universe (in case our universe is
closed), the compact Friedman-Robertson-Walker model. The mixmaster
metric exhibits very complicated dynamical behaviour near its curvature
singularities, illustrating an interesting non-linear aspect of the
Einstein equations. For a recent discussion of the characterization of
chaos in general relativity and the mixmaster toy-model gravitational
collapse, see e.g. S.E.Rugh \cite{SERugh94}.

The governing Einstein equations for the mixmaster metric are given by
a set of three second order ordinary differential equations
\begin{eqnarray} \label{treordinary}
2 \ddot{\alpha} & =  (\e^{2\beta}-\e^{2\gamma})^2 - \e^{4\alpha}
\nonumber \\
2 \ddot{\beta} & =  (\e^{2\gamma}-\e^{2\alpha})^2 - \e^{4\beta} \\
2 \ddot{\gamma} & =  (\e^{2\alpha}-\e^{2\beta})^2 - \e^{4\gamma},
\nonumber
\end{eqnarray}
where $\alpha$, $\beta$ and $\gamma$ are the so-called scale factors
of the metric, and a dot denotes derivative with respect to the
(logarithmic) time variable $\tau$, see Landau and Lifshitz \cite{MTWLL}
\S 118. The equations (\ref{treordinary}) admit a first integral
\begin{equation}
\fl I=4(\dot{\alpha} \dot{\beta} + \dot{\beta} \dot{\gamma} +
\dot{\gamma} \dot{\alpha})
-\e^{4\alpha}-\e^{4\beta}-\e^{4\gamma}
+2(\e^{2(\alpha+\beta)}+\e^{2(\beta+\gamma)}+\e^{2(\gamma+\alpha)}) .
\end{equation}
The {\em mixmaster universe}, with its interesting oscillatory behaviour
of the scale functions near its space-time singularity, arises exactly
at $I=0$. The cosmological model was first analyzed by V.A. Belinski
and I.M. Khalatnikov \cite{Belinski69} and independently by C.W. Misner
\cite{Misner69}, who showed that the asymptotic evolution could be
accurately described as a simple combinatorial evolution of the axes
governed by the Gauss map (see also Barrow \cite{Barrow} and Mayer
\cite{Mayer}). The appearance of the Gauss map may seem to indicate
chaotic behaviour of the system but, as the transformation to the
relevant coordinates is singular and maps non-closed orbits into
closed orbits, it is not clear to what extent conclusions regarding the
inherent chaos of the Gauss map can be carried back to the analysis of
the differential system governing the mixmaster universe.

Following the notation of Contopoulos et al. \cite{Contopoulos}, we
first perform a change of variables~:
\begin{equation}
\begin{array}{lll}
 x=2\alpha & y=2\beta & z=2\gamma  \\
 p_x =-(\dot{y} +\dot{z}) \qquad & p_y =-(\dot{z}+\dot{x}) \qquad &
p_z =-(\dot{x}+\dot{y})
\end{array}
\end{equation}
and find that the system can also be described as governed by the
Hamiltonian
\begin{equation}
\fl H=\frac{1}{4}(p^{2}_{x}+p^{2}_{y}+p^{2}_{z}-2p_x p_y -
2p_y p_z -2p_z p_x )
+\e^{2x}+\e^{2y}+\e^{2z}-2\e^{x+y}-2\e^{y+z}-2\e^{z+x} \equiv -I .
\label{eq:Hamil}
\end{equation}
For the purpose of the Painlev\'{e} analysis a further change of
variables is made
\begin{equation}
X=\e^x \qquad Y=\e^y \qquad Z=\e^z ,
\end{equation}
and we finally end up with this version of differential equations:
\begin{equation}
\fl
\begin{array}{lll}
   2 \dot{X} = X (p_x-p_y-p_z) \quad &
   2 \dot{Y} = Y (p_y-p_z-p_x) \quad &
   2 \dot{Z} = Z (p_z-p_x-p_y) \\
   \dot{p}_x = 2 X (Y + Z - X) &
   \dot{p}_y = 2 Y (Z + X - Y) &
   \dot{p}_z = 2 Z (X + Y - Z) ,
\end{array}
 \label{eq:BianchiIX}
\end{equation}
in which the first integral takes the form
\begin{equation}
\fl  I = -\frac{1}{2} ( p_x^2+p_y^2+p_z^2) + \frac{1}{4}(p_x+p_y+p_z)^2
    - 2(X^2+Y^2+Z^2) + (X+Y+Z)^2 .
  \label{eq:I}
\end{equation}

The sign of the first integral characterizes three different phases of
the model. This can be seen from the scale invariance of the
differential equations (\ref{eq:BianchiIX}) under the transformation
\begin{eqnarray}
  X \r c X  \qquad &  Y \r c Y  \qquad &  Z \r c Z \nonumber \\
  p_x \r c p_x  &  p_y \r c p_y  &  p_z \r c p_z \\
  I \r c^2 I  & \tau \r c^{-1} \tau & \nonumber
\end{eqnarray}

In popular terms we may say that $I=0$ characterizes a phase transition
for the system, as the qualitative description of the time evolution
depends strongly on the sign of $I$ \cite{RughM}. In particular we note
here that since the time variable scales with $|I|^{-1/2}$, the Lyapunov
exponent for the system will have to vanish for $I\r 0$, revealing an
apparent non-chaotic behaviour at $I=0$, i.e. for the mixmaster
universe. Due to this scale invariance we are free to choose an
arbitrary value of $I$ when we look for periodic orbits of the
mixmaster system for $I<0$, and we will choose the value $I=-1$.

\newcommand{\g}{\gamma}

\section{Painlev\'{e} analysis.}

The Painlev\'{e} analysis, when applied to (\ref{eq:BianchiIX}), showed
\cite{Contopoulos} that the system passes the (ARS) Painlev\'{e} test
for the case $I=0$ and therefore, according to the cited conjecture, that
the system is probably integrable. The analysis does not, however,
depend on the value of $I$. Here we focus on one of the singular
solutions (cf. p. 5798 in Contopoulos et al. \cite{Contopoulos}):

Inserting and identifying coefficients in the Laurent series expansion
show that the singular expansions
\begin{equation}
\begin{array}{lll}
   X =  \frac{\i}{s} + \g_1 s + \dots \quad &  Y =  x_2 s + \dots
     \quad  & Z =  x_3 s + \dots \\
   p_x = -  \frac{2}{s} + b + c s + \dots &
         p_y = p_2 +   2\i x_2 s + \dots & p_z = p_3 +   2\i x_3 s + \dots
\end{array}
\end{equation}
satisfy the set of equations (\ref{eq:BianchiIX})  to zero'th order
in $s = t-\tau_0$, provided $b=p_2+p_3$ and $c=2\i [x_2+x_3-2\gamma_1]$.
Here, $t$ is the complex time variable and $\tau_0, \g_1, x_2, x_3,
p_2$ and $p_3$ are the six free parameters in the generic Painlev\'{e}
expansion. Computing $I$ to zero'th order
(all other terms must vanish due to the time invariance of $I$) we get :
\begin{equation}
 I = p_2 p_3 + 2 \i (x_2 + x_3 - 3 \g_1) \ \ ,
\end{equation}
which is a constant of motion  but in general {\em does not} vanish.
Thus we see, as is also the case for the other classes of singular
solutions in \cite{Contopoulos}, that the set of equations has the
(ARS) Painlev\'{e} property independent of the value of the first
integral $I$.

\section{Closed orbits and non-integrability.}

One way to characterize non-integrability is by means of non-zero
Lyapunov exponents. In general, each first integral of a Hamiltonian
system implies the vanishing of two Lyapunov exponents due to
preservation of symplectic two-forms \cite{HDMeyer}. In practice this
characterization is not very useful. First, it is not easy to establish
numerically due to the infinite time limit involved, and second,
because the reverse need not be true, i.e. the vanishing of 2$n$
Lyapunov exponents does not imply the existence of $n$ first integrals.
However, in general the vanishing of Lyapunov exponents occurs for
almost all trajectories of infinite length. In particular, we shall
consider {\em closed orbits}, i.e. trajectories which return to a given
point, $x$, in the phase space after some period  $T$. Periodic orbits
have the great advantage (see also \cite{Cvitanovic}) that they are
completely determined by their finite time behaviour and therefore
provide a numerically easy and reliable way of stating
non-integrability of the system.

We write the differential equations
(\ref{eq:Hamil})-(\ref{eq:BianchiIX}) in the form :
\begin{equation}
\frac{\d}{\d t} x = \left(
    \begin{array}{rr}
       \BFACE{0}  & -\BFACE{1} \\
       \BFACE{1} & \BFACE{0}
    \end{array}
 \right) \nabla H(x)  \equiv J \nabla H (x)
       \ \ , \ \ \ x = (p,q) \in R^d \times R^d
\end{equation}
(here $d=3$) and we denote by $\phi^t_H(x)$ the flow obtained by
integrating this equation. A real-analytic function $I$ on $R^{2d}$ is
said to be a first integral if it commutes with $H$, i.e. if their
poisson bracket vanishes~:
\begin{equation}
       \{I,H\} \equiv (\nabla I, J \nabla H) \equiv 0
\end{equation}
$(\cdot,\cdot)$ being the usual inner product in $R^{2d}$. In
particular \cite{Arnold},  this condition implies that the flow
$\phi^s_I(x)$, generated by $J\nabla I$, commutes with $\phi^t_H$. Thus,
we have the two relations~:
\begin{equation}
\begin{array}{lll}
  I \circ \phi^t_H  = I & t \in R \\
  \phi^s_I \circ \phi^t_H = \phi^t_H \circ \phi^s_I \qquad & s,t \in R
\end{array}
\end{equation}
Assume now that we have found a periodic orbit $x=\phi^T_H(x)$ of our
Hamiltonian flow with return time $T>0$. We may use this periodicity
condition in our two relations above, taking derivatives in $x$ and in
$s$ (at $s=0$), respectively, to obtain (tr meaning transpose)~:
\begin{equation}
\eqalign{\nabla I (x)^{\tr}  D\phi^T_H(x)  = \nabla I(x)^{\tr} \ , \\
      J \nabla I (x)    = D\phi^T_H(x)   J \nabla I(x)  \ . }
\end{equation}
We can replace $I$ with $H$ or any other first integral (provided they
all commute) and conclude that the (transposed) set of $\nabla H$,
$\nabla I$, etc. generates a set of left invariant eigenvectors of
$D\phi^T_H(x)$ while by multiplying by $J$ we obtain a set of right
invariant eigenvectors of the same matrix.

The two sets are evidently mutually orthogonal (as the first integrals
commute). Thus we may orthonormalize each set and take their union to
be a basis of $R^{2d}$. A straightforward calculation shows that in
this basis $D\phi^T_H(x)$ takes the form of an upper (or lower)
triangular matrix with $1$ in the diagonal. Thus, we conclude that if
the system has $d$ independent first integrals at a periodic orbit then
the stability matrix has only one eigenvalue, 1, with degeneracy $2N$.

As Poincar\'{e} already noted \cite{Poincare}, the presence of non-unit
eigenvalues does not exclude integrability but could germinate from a
degeneracy of first integrals.
The simplest example of this is the stiff pendulum which, when placed
at rest in its top position, is in an unstable fixed point. The
separatrices \cite{Arnold} through this point separate the phase
space into oscillatory modes and two types of rotational modes. The
eigenvalues of the stability matrix for the fixed point for any
non-zero time are not on the unit-circle. The arguments above do not
apply to this case as we here have $\nabla H = 0$ (a
special case of degenerating first integrals). More complicated
situations arise in integrable systems in higher dimensions when,
for closed orbits, the gradients of the first integrals become
linearly dependent.

As mentioned above, a rigorous proof would require the presence of
a transversal homoclinic intersection \cite{Cushman,Niko84}.
However, we insist that an abundance of unstable periodic orbits
strongly indicates non-integrability of the dynamical system, i.e. that
it cannot possess $d$ commuting analytic first integrals which are
independent almost everywhere.

\section{Unstable periodic orbits for $I < 0$.}

Finding periodic orbits in a low-dimensional dynamical system such as
(\ref{treordinary}) means solving the equation
\begin{equation}
\phi^T (x) = x
\label{obvious}
\end{equation}
and is a straightforward, if tedious, task. One starts with random
initial conditions and lets the system evolve until a suitable
close-return occurs. What {\em is} a suitable close-return is as much
determined by elapsed time as distance between initial and final values
of the variables, due to the high instability of the dynamics. One then
uses the initial conditions of the close return trajectory as a starting
guess in an extended Newton search for a solution to (\ref{obvious}).
This takes the following steps: \\
1. Choose a Poincar{\'e} section, i.e. a $(2N-1)$-dimensional surface
that the trajectory will cross. This is rather arbitrary, but a plane
through the initial point perpendicular to the initial velocity works
well. \\
2. Integrate the equations of motion including the Jacobian matrix
$M$ (here $v=J\nabla H$)
\begin{equation}
\dot{M}=\frac{\partial v(x(t))}{\partial x} M, \qquad M_{t=0} = 1 ,
\end{equation}
until the trajectory crosses the Poincar{\'e} section at $x_f$, close
to the initial point, $x_i$. \\
3. Assume that the trajectory is close to a periodic orbit and solve
the linearized version of (\ref{obvious}):
\begin{equation}
x  =  \phi^T (x) \approx \phi^T(x_i )+M(x-x_i ) = x_f + M(x-x_i),
\end{equation}
i.e. solve
\begin{equation}
(1-M)x  =  x_f - Mx_i .
\label{linobv}
\end{equation}
This presents the problem that since $M$ has 1 as (at least) a double
eigenvalue, $1-M$ is not invertible. To remedy this, we augment the
equation (\ref{linobv}) with 2 more equations and 2 more variables
\cite{Freddy}. The two extra equations serve the purpose of keeping the
solution $x$ on the Poincar{\'e} section and preserving the first
integral $I$ during the Newton iteration, while the two variables add
corrections to $x$ along the right/left eigenvectors of the eigenvalue
1, i.e. $v(x)$ and $\nabla I$, in order to insure the existence of a
solution to (\ref{linobv}). \\
4. Repeat until satisfactory convergence is obtained.

We plot two examples of periodic orbits in figures \ref{simple} and
\ref{complex}. In tables \ref{cycles} and \ref{tabel2} we give the
initial conditions, periods and eigenvalues of a few of the periodic
orbits found through this method. As can be seen, the eigenvalue 1 has
multiplicity 2 in all our examples. We have found approximately 200
unstable periodic orbits, of which around 50 were found in a random
search through phase space, whereas the remaining number was found in
a systematic search for members of infinite families of orbits as
described below. This numerical result makes it highly unlikely that
the system is integrable for $I<0$.

The existence of unstable periodic orbits does not directly lead to
chaos in the sense of a positive Lyapunov exponent in the mixmaster
system for $I<0$. In fact, we have been able to find series of periodic
orbits that indicate that the Lyapunov exponent will converge to zero
for infinite time. Consider a family of periodic orbits where one of
the scale factors, $\alpha$, makes one oscillation while the other two,
$\beta$ and $\gamma$ make $n$ oscillations. In figure \ref{10cyc} such
an orbit with $n=10$ is shown. We find that for such a family the
largest eigenvalue of $M$, $\Lambda_n$ grows linearly with $n$, whereas
the period, $T_n$ grows as $\sqrt{n}$. This means that for the
instability exponent $\lambda_n$, i.e. the local Lyapunov exponent of
the individual periodic orbits, we have:
\begin{equation}
\lambda_n \equiv \frac{\log \Lambda_n}{T_n} \sim
\frac{\log n}{\sqrt{n}} ,
\end{equation}
which will tend to zero for $n$ going to infinity. This is shown in
table \ref{ncyctab} and figure \ref{ncycfig}. The limit of such a
family of periodic orbits is therefore marginally stable. In the
computation of the Lyapunov exponent, marginally stable regions seem to
dominate making the exponent converge to zero. This conclusion is
supported by a direct computation of the finite time approximation of
the Lyapunov exponent. Starting with 1000 random initial conditions, we
have computed the average finite time Lyapunov exponent, which appears
to fall off as a power law with increasing $\tau$. This result is shown in
figure \ref{lyapexp}.

We therefore conclude that the mixmaster system (\ref{treordinary})
for $I<0$ is neither integrable nor chaotic in the sense of having a
positive maximal Lyapunov exponent.

\section{Concluding remarks}

We have noted that, according to the recent analysis by Contopoulos et
al. \cite{Contopoulos}, the mixmaster cosmological model passes the
Painlev\'{e} test not only for the case $I=0$, i.e. the vacuum Einstein
equations for which the behaviour of the axes is described by the
BKL-combinatorial model, but for arbitrary values of the first integral
$I$. At the same time we have shown that when $I<0$, the model is
probably non-integrable through the existence of an abundance of
unstable periodic orbits.

Thus, the model provides a first example of a Hamiltonian system of
non-linear differential equations that passes the ARS-test for the
Painlev\'{e} property but which is not integrable. One may still speculate
whether the mixmaster model has the full Painlev\'{e} property (the answer
appears to be negative according to the recent results of Contopoulos et
al. \cite{Contopoulos94ny} and those  of Latifi et al. \cite{Latifi94}),
or whether the constraint $I=0$ may be incorporated in a dynamical
context or in the Painlev\'{e} analysis in some way, avoiding the
non-integrable periodic solutions described in this article and
rendering the system integrable.

\ack
F. Christiansen is supported by the  European Union Human Capital and
Mobility grant no. ERB4001GT921004 and is grateful for the hospitality
of INO, Firenze. S.E.Rugh would like to thank support from the Danish
Natural Science Research Council grant no. 11-8705-1. H.H. Rugh is
supported by the European Human Capital and Mobility grant no.
ERB CHB GCT 92006 and IHES.

We thank for fruitful discussions with Arne Lykke Larsen and with
A. Ramani and B. Grammaticos at the Ecole Polytechnique and
G. Contopoulos for pointing out a flaw in a draft of the paper.

\begin{table}
\caption[tab1]{Initial conditions for some periodic orbits.
Entries i, v and vi correspond to figures
\ref{simple}, \ref{complex}+\ref{complexb} and \ref{10cyc}.
The number of digits given corresponds to the
accuracy of the numerical methods employed as estimated by using 2
different integration routines and 2 different tolerance levels in the
routines. \label{cycles} }
\lineup
\begin{indented}
\item[]\begin{tabular}{@{}lllll}
\br
 & \centre{3}{Initial values} & Period  \\ \mr
 $\#$ & $\alpha$, $\dot{\alpha}$ & $\beta$, $\dot{\beta}$ &
 $\gamma$, $\dot{\gamma}$ & $T$ \\ \mr
i & \-0.7465064136 & \-0.04110593028 & \-0.31459136109 &\05.6040350442 \\
 & 0.0 & 0.57358291296 & \-0.6330334642 & \\ \mr
ii & 0.28088191655 & \-1.4043414748 & \-0.21644617646  & \07.9472870699 \\
 & 0.0 &  \-0.19132448256 & 0.084543111864 & \\ \mr
iii & \-0.85822654286  & 0.296817610585 & \-0.374109406546 & \08.54414096178 \\
 & 0.0 & 0.0 & 0.0 & \\ \mr
iv & 0.399801735 & \-0.141272123 & \-0.26896997 & 13.4135432128 \\
 & 0.0 & 1.044336138 & \-0.47263976 & \\ \mr
v & 0.13651016517 & \-0.83335865226 & \-0.5987346358 & 14.6263998708 \\
 & 0.0 & 0.424450201779 & \-0.32431285874 & \\ \mr
vi & 0.67215626848 & \-0.04851348437 & \-0.04851348437 & 16.15030271 \\
&   0.0 & 0.2302667762 & \-0.230266776 & \\ \mr
vii & \-1.0663696240 & \-2.7978105880 & 0.07780069395 & 16.435444052 \\
 & 0.0 &  0.0668561349 & 0.3463999832 & \\ \br
\end{tabular}
\end{indented}
\end{table}
\begin{table}
\caption[tab2]{Eigenvalues for the periodic orbits of Table \ref{cycles}.
The lesser accuracy of the unit eigenvalues is caused by
the fact that they appear as double eigenvalues.
\label{tabel2} }
\lineup
\begin{indented}
\item[]\begin{tabular}{@{}llll}
\br
 & \centre{3}{Eigenvalues} \\ \mr
i & \0\0\-29.654128172 & 1.00000 & \-0.948098014\0\0 + $i$0.3179782317 \\
  & \0\0\0\-0.033722117682 & 1.00000 & \-0.948098014\0\0 $-$ $i$0.3179782317
 \\ \mr
ii & \0\-103.545671685 & 1.00000\0\0\0\0\0 & \-0.69755560723 + $i$0.71653065169
\\
  & \0\0\0\-9.657574128 $10^{-3}$ & 1.00000 &
\-0.69755560723 $-$  $i$0.71653065169 \\ \mr
iii & \0136.451552742 & 1.00000 & \-0.48706495012 + $i$0.87336575062 \\
  & \0\0\07.328608432 $10^{-3}$ & 1.00000 &
\-0.48706495012 $-$  $i$0.87336575062 \\ \mr
iv & 2540.8960825 & 1.00000 & 1.044452134\0\0 \\
  & \0\0\03.9356194\0\0 $10^{-4}$ & 1.00000 & 0.957439758\0\0 \\ \mr
v & \-5843.077705 & 1.0000 & 0.5307158319\0 +  $i$0.8475498249 \\
 & \0\0\0\-1.7114268\0\0 $10^{-4}$ & 1.0000  &
0.5307158319\0 $-$ $i$0.8475498249 \\ \mr
vi & \0\-643.8934046 & 1.00000 & 0.0021357407\0 +  $i$0.9999977193 \\
 & \0\0\0\-1.55305209\0 $10^{-3}$ & 1.00000 &
0.0021357407\0 $-$ $i$0.9999977193 \\ \mr
vii & 1926.43285564 & 1.0000  & 2.170605581\0\0 \\
 & \0\0\05.1909413\0\0 $10^{-4}$ & 1.0000 & 0.4607009253\0 \\ \br
\end{tabular}
\end{indented}
\end{table}

\begin{table}
\caption[tab3]{Largest stability eigenvalue, period and stability
exponent for the (1,$n$,$n$) family of periodic orbits. $T_n$ and
$\Lambda_n$ vs. $n$ are plotted in figure \ref{ncycfig}.
Beyond $n\approx 40$ it gets increasingly
difficult to follow the trajectories numerically because the dynamics
is determined by the decreasing difference between the two
oscillating scale factors $\beta$ and $\gamma$.\label{ncyctab} }
\lineup
\begin{indented}
\item[]\begin{tabular}{@{}llll}
\br
$n$ & $T$ & $\Lambda$ & $\lambda \equiv \log (|\Lambda |)/T$ \\ \mr
1 & \05.6040350441 & \0\0\-29.654128172 & \-0.11430453548 \\
2 & \07.9472870699 & \0\-103.5456716852 & \-0.04481126745499 \\
3 & \09.42569727483 &  \0\-186.757808548 & \-0.0280031805718 \\
5 & 11.7015648632 & \0\-331.02446237 & \-0.0175279864076 \\
10 & 16.150302715 & \0\-643.89340465 & \-0.0100444159622 \\
15 & 19.649946461 & \0\-967.1141240 & \-0.0071080716705 \\
20 & 22.618966753 & \-1292.2369183 & \-0.0055439756742 \\
25 & 25.242560127 & \-1617.915587 & \-0.00456692177 \\
30 & 27.61864459 & \-1943.8400780 & \-0.00389559861571 \\
40 & 31.84403793 & \-2596.030034 & \-0.003028369681 \\
46 & 34.12932414 & \-2987.456574 & \-0.002678592127 \\ \hline
\end{tabular}
\end{indented}
\end{table}

\newpage

\section*{Figure Captions}

\begin{figure}[h]
\caption[figsimple]{The simplest periodic orbit of the mixmaster
system corresponding to entry i in table \ref{cycles}. The 3 scale
factors $\alpha$, $\beta$ and $\gamma$ each make one identical
oscillation. The trajectory takes the form of a slightly
distorted circle.}
\label{simple}

\caption[figcomplex]{A more complex periodic orbit of the mixmaster
system corresponding to entry v in table \ref{cycles}. The
largest eigenvalue $\Lambda$ of the Jacobian for this trajectory is
$\approx$ 5800 for a period $T$ of only 14.62. }
\label{complex}

\caption{The same trajectory as above but shown projected on the
$(\alpha,\beta )$-plane.\mbox{\hspace{1.5cm}}}
\label{complexb}

\caption[fig10cyc]{An example of a periodic orbit in the family
with respectivelt 1, $n$ and $n$ oscillations of the three scale
factors with $n=10$. The initial values correspond to entry vi in
table \ref{cycles}.
For increasing $n$ the difference between the two oscillating scale
factors becomes smaller and smaller.}
\label{10cyc}

\caption[figncyc]{The power law dependence of the largest instability
eigenvalue $\Lambda$ of the Jacobian, $M$, and the period, $T$, vs.
$n$ in the (1,$n$,$n$) family of periodic orbits. We have found
orbits up to $n=46$. For $n\geq 6$ we estimate the slope to
be .495 for $T$ and 1.006 for $\Lambda$. Some of the data points
are listed in table \ref{ncyctab}.}
\label{ncycfig}

\caption{The finite time Lyapunov exponent $\lambda_{\tau}$ computed
as an average of 1000 random trajectories. For numerical reasons we
use the largest eigenvalue of $M^{\tr}M$. $\tau=70$ is approximately
the time limit, where we lose information about the {\em lowest}
eigenvalue due to numerical underflow. We believe this explains the
bend of the curve near this value and therefore that the power law
fall-off holds even for $\tau > 70$. The power law fall-off
indicates that the infinite time Lyapunov exponent is 0.}
\label{lyapexp}
\end{figure}

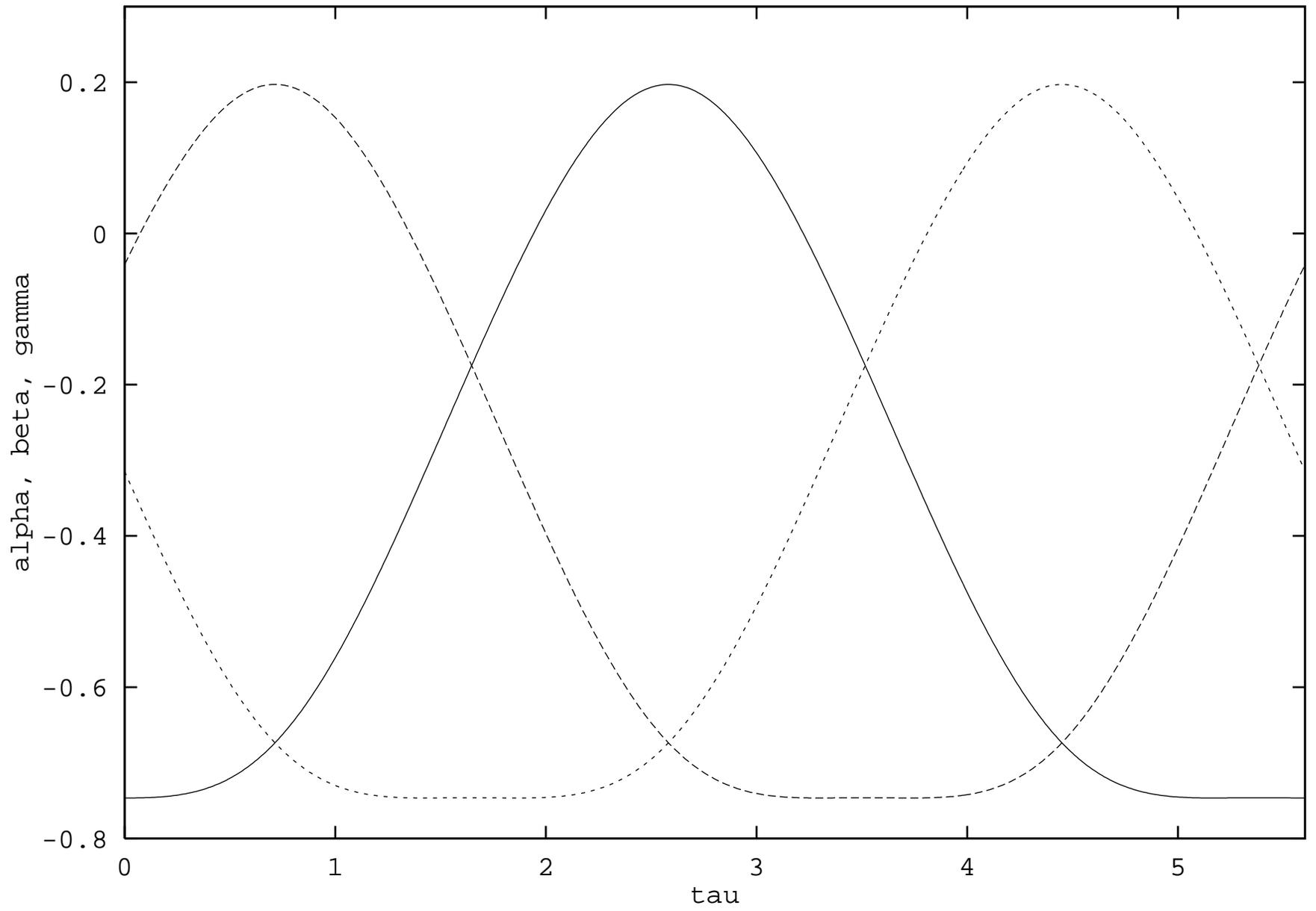
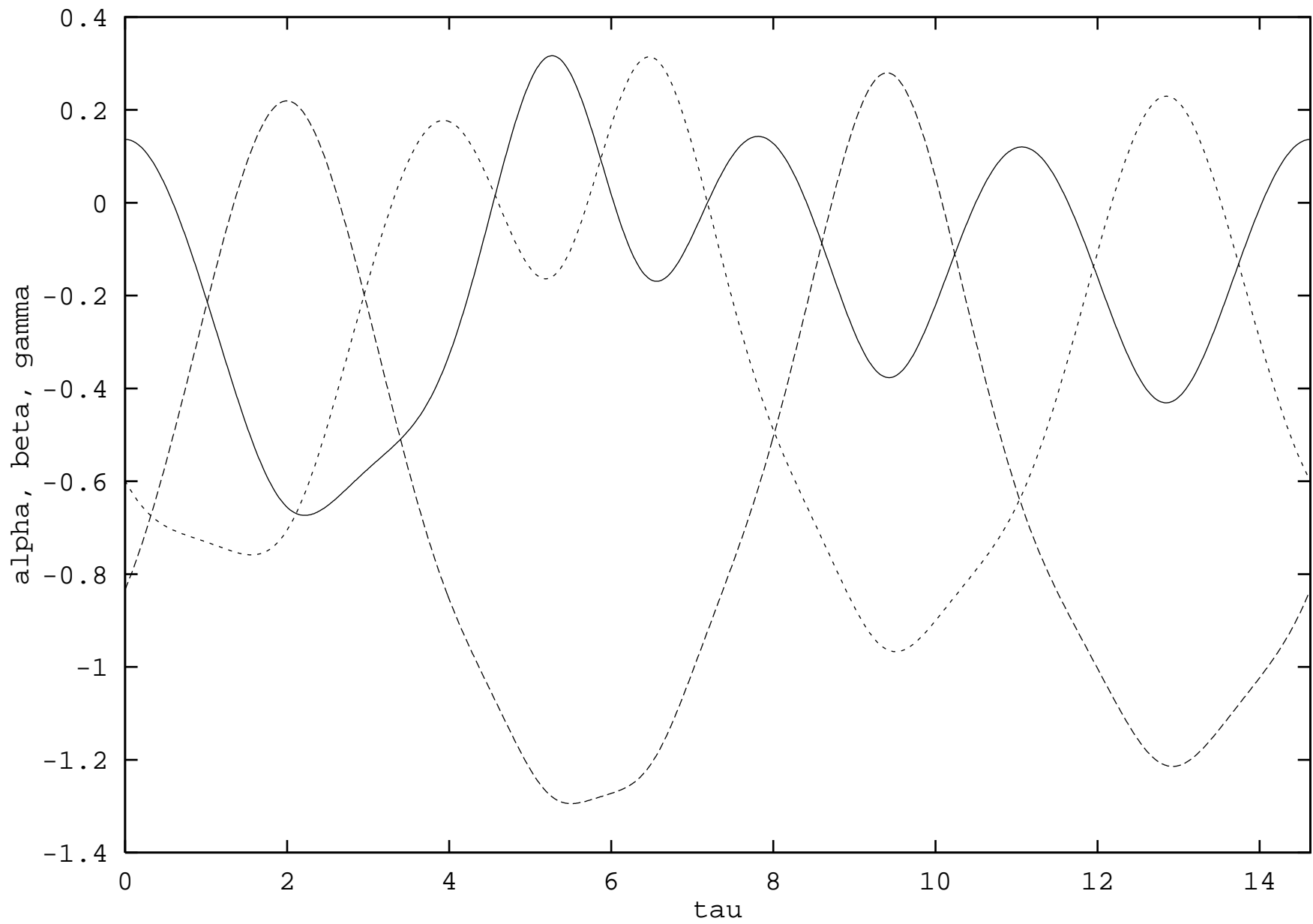
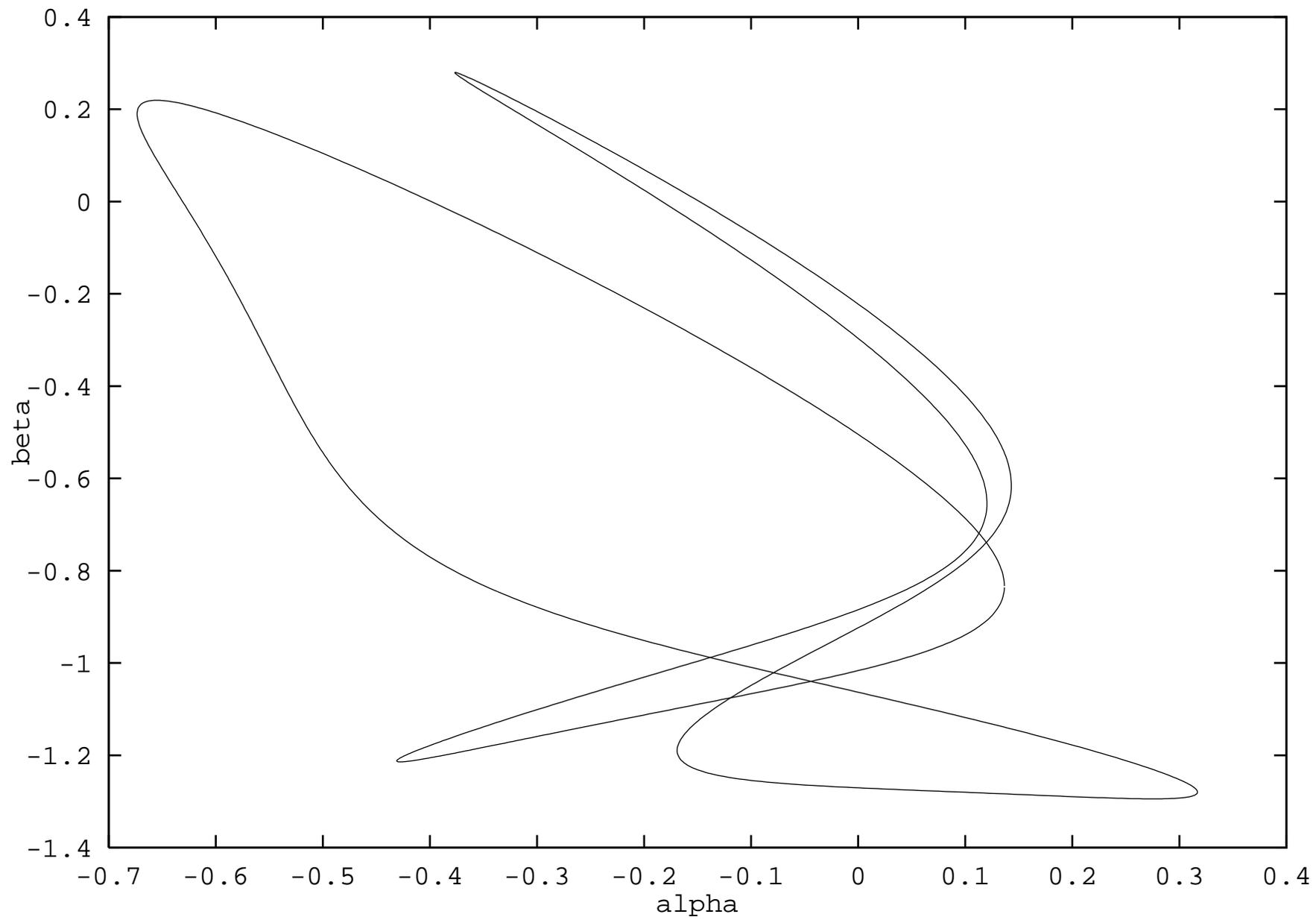
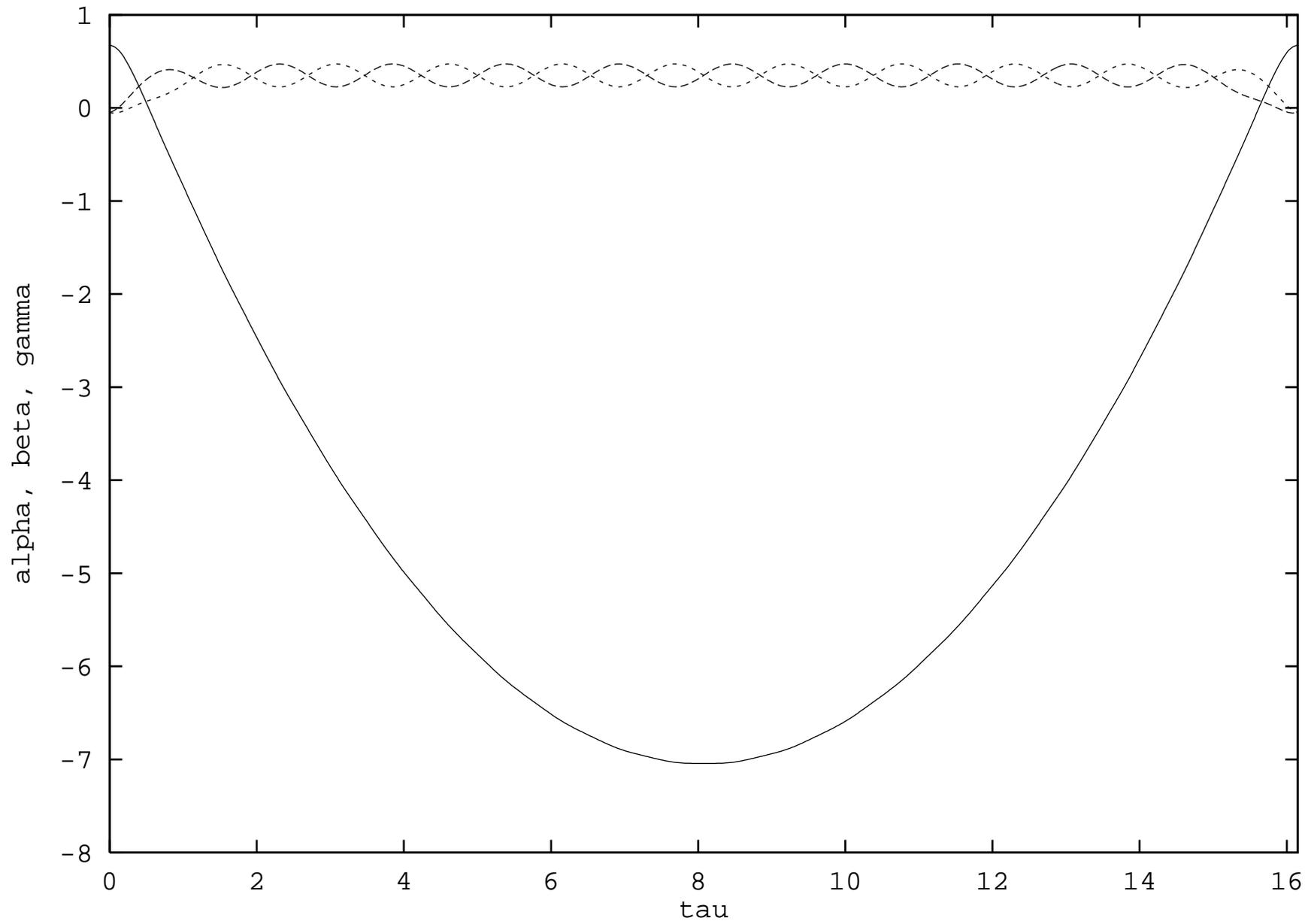
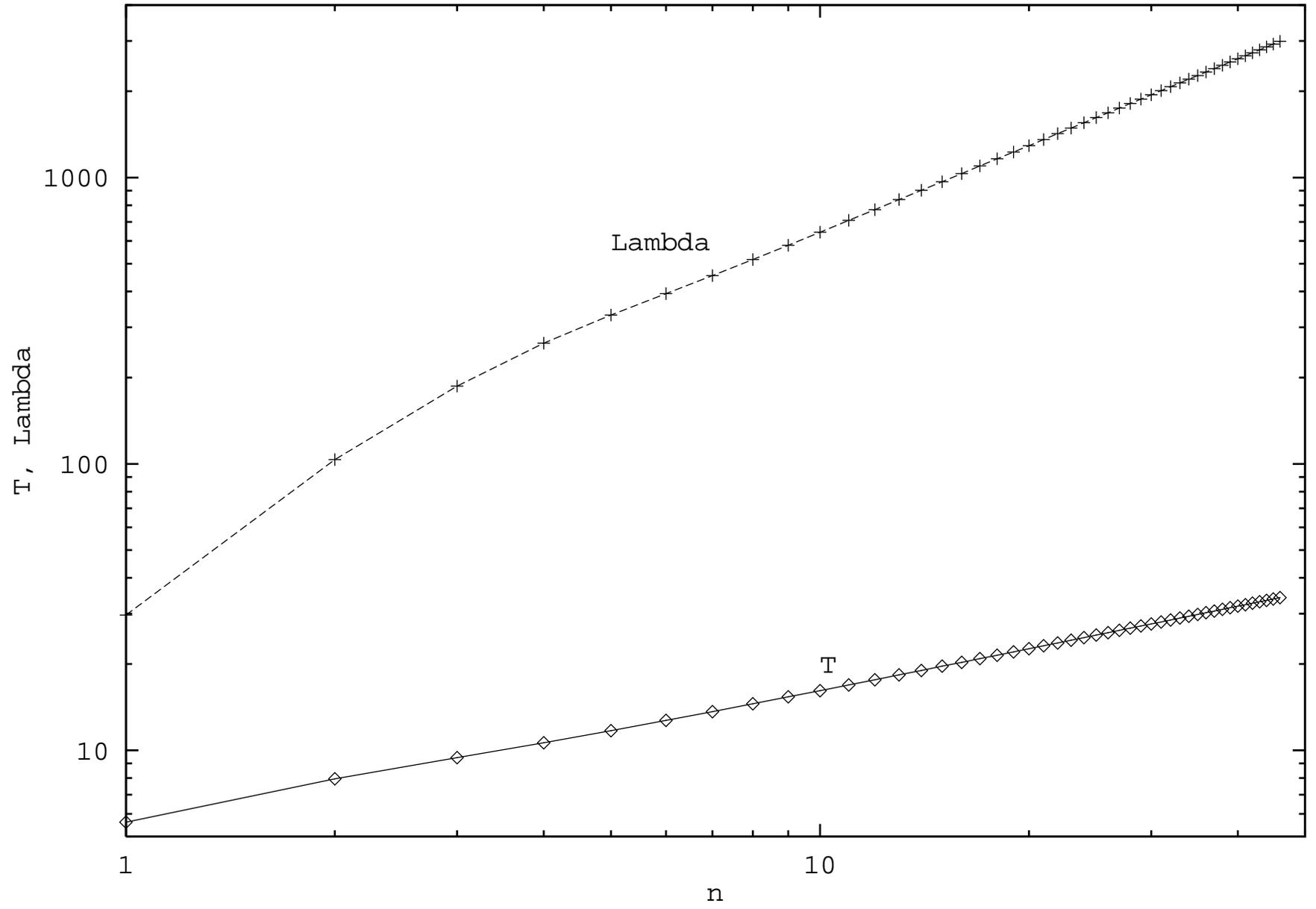
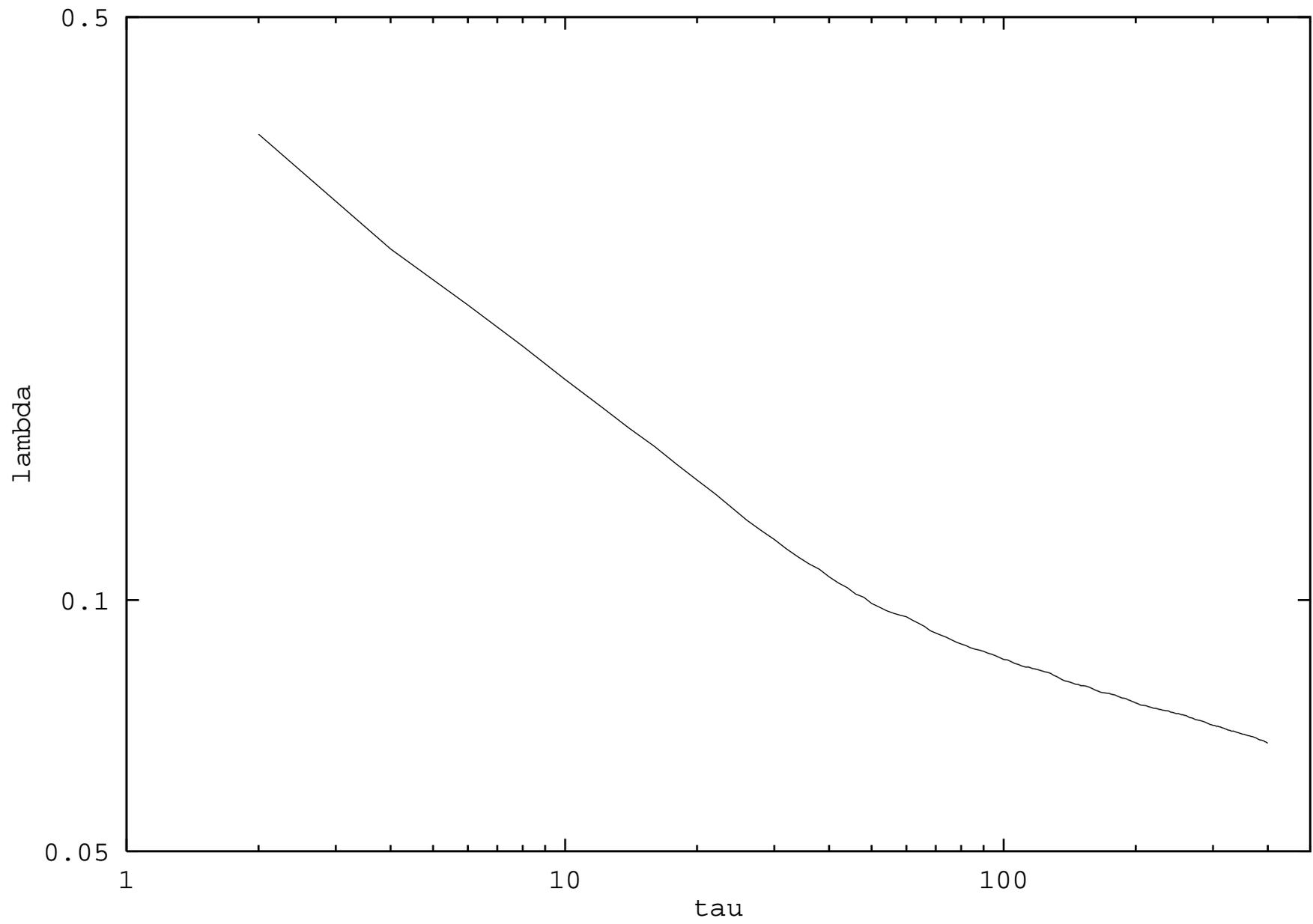
\end{document}